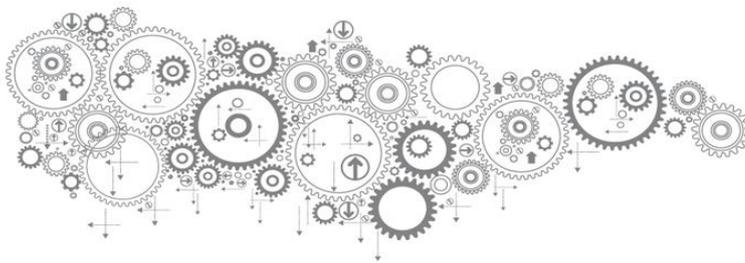

# Gimbal Actuator Modeling for a Spin-Stabilized Spacecraft Equipped with a 1DoF Gimbaled-Thruster and two Reaction Wheels

**Hamed Kouhi[1,*], Mansour Kabganian[2], Farhad Fani Saberi[3], Fatemeh Ghorbani[4]**

[1] Assistant Professor, Department of Mechanical Engineering, University of Guilan, Rasht, Iran.
[2] Professor, Department of Mechanical Engineering, Amirkabir University of Technology, Tehran, Iran.
[3] Assistant Professor, Space Science and Technology Institute, Amirkabir University of Technology, Tehran, Iran.
[4] Faculty of Electrical Engineering, K. N. Toosi University of Technology, Tehran, Iran.

## ABSTRACT

Attitude control of spacecraft during an impulsive orbital maneuver is a vital task. Many spacecraft and launchers use the gimbaled thrust vector control (TVC) in their attitude control system during an orbital maneuver. Mathematical modeling of the gimbal actuator is an important task because we should show the applicability of the gimbaled-TVC in a spacecraft. In this paper, a spin-stabilized spacecraft equipped with one degree of freedom (DoF) gimbaled-thruster and two reaction wheels (RWs) is considered. The control goals are disturbance rejection and thrust vector (spin-axis) stabilization based on one DoF gimbal actuator and two RWs. The gimbal is assumed to be equipped with a gearbox and a DC electric motor. This actuator must supply the gimbal torque to rotate the spacecraft nozzle. The mathematical model of the mentioned spacecraft is extended with respect to the DC motor equations. In order to investigate the applicability of the proposed method, an industrial DC electric motor is considered for the gimbal actuator. The simulation results prove that an industrial DC electric motor is able to be used for attitude control of the mentioned spacecraft. The simulation results indicate the applicability of the proposed control method in an impulsive orbital maneuver.

## 1. INTRODUCTION

In an impulsive orbital maneuver, thrust vector offset from the center of mass generates a large disturbance torque that results in a thrust vector deviation from the desired inertial direction (C.M), (see [1]). As a result, a powerful attitude control system is needed to reject the mentioned disturbance torque [2].

TVC method works by directing the thrust vector through the spacecraft C.M. This robust scheme with many advantages, works with an electrical actuator without fuel consumption. In this method, an active control torque can easily reject the disturbance created by the thrust vector misalignment [3]. When the disturbance torque amount is so larger than the attitude control level, a gimbaled-TVC is applicable (Apollo, Cassini [4], [5], and launchers). gimbaled-TVC is simpler and more efficient than the other TVC methods such as moving plate (see [6]). The gimbaled-thruster enables us to reduce the requirements of the ground C.M positioning and simplify the attitude control system [7], [8], [9]. In liquid propellant rockets, the dynamical interaction between the rotatable nozzle and spacecraft body is very small (see [1], [10-12]).

However, for a spacecraft equipped with a gimbaled solid rocket motor (SRM), the dynamical interaction makes a nonlinear two-body dynamics [2], [8], [9]. Note that, in the case of gimbaled-TVC, a high level control torque is essential to control the gimbal angle. In this paper, a DC electric motor is considered as the gimbal actuator. In order to investigate the applicability of the thrusting maneuver based on gimbaled-TVC, the ability of the actuator must be checked; therefore, mathematical modeling and analysis should be carried out. In many studies in which gimbaled-thruster is applied (such as [2] and [9]) the gimbal actuator modeling is not addressed.

---

[*] Corresponding Author, Email: hamed.koohi@guilan.ac.ir

In work [2], the gimbal actuator has an important and vital role in disturbance rejection and thrust vector stabilization. Thus, mathematical modeling is considered for the gimbal actuator in our work. We have also employed a DC electric motor with a gearbox. This study aims to investigate some industrial DC motors and indicate that by using them, the spacecraft can be controlled during a thrusting maneuver. The mentioned spin-stabilized spacecraft includes a 1DoF gimbaled-thruster and two RWs.

## 2. DYNAMIC MODELING

The dynamics equations of a spin-stabilized spacecraft equipped with a 1DoF gimbaled-thruster and two RWs is derived in [2]. In this reference, the gimbal actuator modeling has not been addressed. In this section a DC electric motor is considered for the gimbal actuator and then based on this assumption the state space model is extended.

### 2.1 Nonlinear Dynamic Modeling

The spacecraft is shown in Figure 1, where the thruster (SRM) has a constant thrust $F_T$.

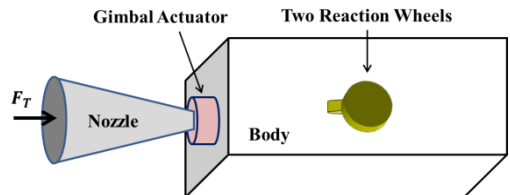

**Figure 1.** A spacecraft equipped with spin-stabilization, two RWs, and a 1DoF gimbaled-thruster

In Figure 2, the free body diagram of the spacecraft body and nozzle is given. Subscripts s, n, o, and T indicate the body, the nozzle, gimbal pivot, and the point of acting the thrust force, respectively. Moreover, $x_s y_s z_s$ and $x_n y_n z_n$ are the body, and the nozzle fixed frame placed in their C.Ms. $\boldsymbol{\tau}_s \in \Re^3$ ($\boldsymbol{\tau}_n$) and $\boldsymbol{\omega}_s \in \Re^3$ ($\boldsymbol{\omega}_n$) are the body (nozzle) external torque and the angular velocity, respectively. $\boldsymbol{M}_o \in \Re^3$ and $\boldsymbol{F}_o \in \Re^3$ are the interaction torque and force at the pivot o. As shown in Figure 3, β is the relative rotation of $x_n y_n z_n$ with respect to $x_s y_s z_s$ using a 1DoF gimbal actuator at the pivot o.

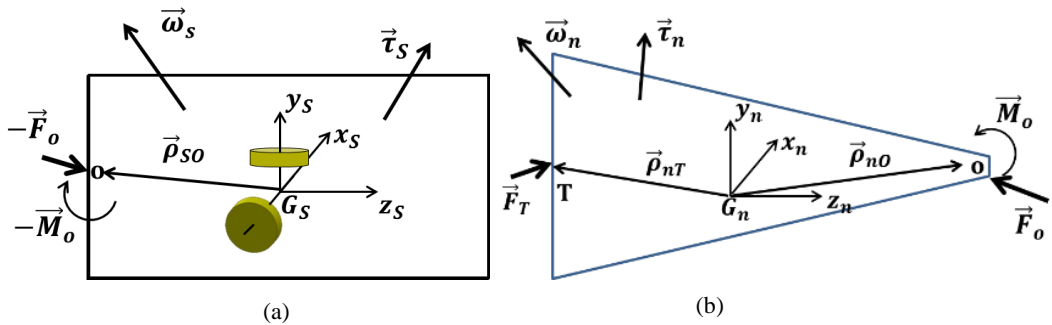

**Figure 2.** Free body diagram of (a) the spacecraft body, and (b) nozzle

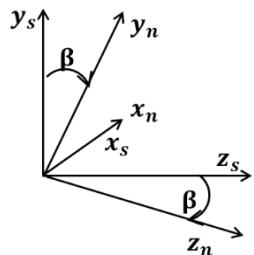

**Figure 3.** Gimbal rotation angle β at the pivot o

Euler momentum equation for the body is:

$$\tau_s + \tau_{Rw} - \rho_{so} \times F_o - M_o = I_s \dot{\omega}_s \qquad (1)$$
$$+ \omega_s \times (I_s \omega_s + H_{Rw}),$$

the dynamics equations modeling is given in [2], in which $\dot{\omega}_s$ as the body angular acceleration is achieved as:

$$\dot{\omega}_s = [I_{ns,T}]^{-1} T_s \qquad (2)$$

where, $T_s = \tau_{ns} + \tau_{Rw} + \rho_T \times F_T - I_r(\dot{\omega}_r + \omega_s \times \omega_r)$
$+ M\rho_{ns} \times (\omega_s \times (\omega_s \times \rho_{so}) - \omega_n \times (\omega_n \times \rho_{no}))$
$\qquad - \omega_s \times (I_s \omega_s + H_{Rw}) - \omega_n \times (I_n \omega_n),$

$\omega_s = \begin{bmatrix} \omega_{sx} & \omega_{sy} & \omega_{sz} \end{bmatrix}^T$, $\omega_n = \omega_s + \omega_r$, $\tau_{ns} = \tau_s + \tau_n$,
$\rho_{ns} = \rho_{no} - \rho_{so}$, $\rho_{sn} = -\rho_{ns}$, $I_{ns} = I_s + I_n$,

$\omega_r = \begin{bmatrix} \dot{\beta} & 0 & 0 \end{bmatrix}^T$, $\dot{\omega}_r = \begin{bmatrix} \ddot{\beta} & 0 & 0 \end{bmatrix}^T$.

And

$H_{Rw} = \begin{bmatrix} h_{r,x} & h_{r,y} & 0 \end{bmatrix}^T$ and $\tau_{Rw} = \begin{bmatrix} \tau_{r,x} & \tau_{r,y} & 0 \end{bmatrix}^T$ represent the RWs angular momentum and axial torques, respectively.

### 2.2 Gimbal Actuator Modeling

In this section, the goal is to find the spacecraft state-space model for the condition that a DC motor is used as the gimbal actuator. In Figure 4, a typical torque-speed curve of DC motor is shown where $V_1$ and $V_2$ represent the two voltage levels. These DC motors have been used in space mechanism as well as in the TVC. It is obvious that by increasing the speed, the torque is decreased linearly.

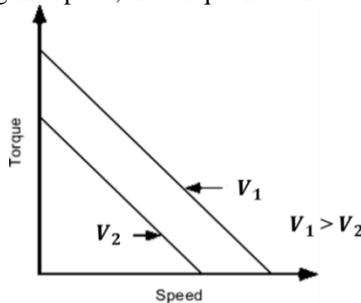

**Figure 4.** Torque-speed curve for a DC motor

In Eq. (**3**), the governed dynamics of a DC motor is given

$$\tau_{em} = k_{2,em}(V_{em} - k_{1,em}\omega_{em}) - J_{em}\dot{\omega}_{em} - \tau_{f,em} \qquad (3)$$

where, subscript *em* stands for the electric motor, $\tau_{em}$ is the output torque, $V_{em}$ is the input voltage, $\omega_{em}$ and $\dot{\omega}_{em}$ are the angular velocity and acceleration of the rotor, $\tau_{f,em}$ represents the friction torque, $k_{1,em}$ and $k_{2,em}$ are the motor constants. Note that $V_{em}$ is selected as the control input.

Motor constants are obtained by the following equation:

$$k_{1,em} = \frac{V_{em,max}}{\omega_{em,max}} \ , \quad k_{2,em} = \frac{\tau_{em,max}}{V_{em,max}} \qquad (4)$$

In which, maximum voltage $V_{em,max}$, stall torque $\tau_{em,max}$, and maximum speed $\omega_{em,max}$ are achieved through the function sheet of each DC motor.

In order to increase the acting torque on the gimbal pivot, a gearbox with the ration of $N_g$ is employed; therefore, the motor's output speed is reduced with the ratio of $1/N_g$. In this paper, the friction torque is considered as viscous friction $\tau_{f,em} = c_{f,em}\omega_{em}$. By the mentioned assumptions, the following relations are established:

$$M_{o,x} = N_g \tau_{em}$$
$$\omega_{em} = \dot{\beta} N_g \tag{5}$$
$$\tau_{f,em} = c_{f,em}\omega_{em} = N_g c_{f,em}\dot{\beta}$$

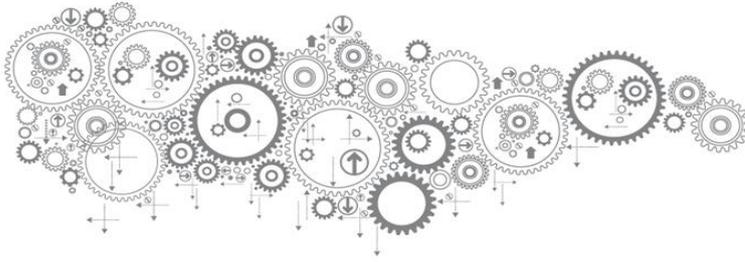

Where, $M_{o,x}$ is the acting torque on the gimbal x-axis.

### 2.3 Spacecraft Kinematics

To obtain the body attitude with respect to the inertial coordinate $X_I Y_I Z_I$, the time derivative of the Euler angles is obtained as:

$$\begin{bmatrix} \dot{\varphi} & \dot{\theta} & \dot{\psi} \end{bmatrix}^T = \boldsymbol{J}_\omega^{Eu}(\varphi, \theta, \psi)\,\boldsymbol{\omega}_s, \tag{6}$$

Where,

$$\boldsymbol{J}_\omega^{Eu} = \begin{bmatrix} 1 & \sin(\varphi)\tan(\theta) & \cos(\varphi)\tan(\theta) \\ 0 & \cos(\varphi) & -\sin(\varphi) \\ 0 & \sin(\varphi)/\cos(\theta) & \cos(\varphi)/\cos(\theta) \end{bmatrix}.$$

The actual velocity change $\Delta v_z$ in the direction of the $Z_I$ is achieved as:

$$\Delta v_z = \int_0^{T_b} a_{\max} \cos(\delta_{FT}(t))\, dt, \tag{7}$$

Where $\delta_{FT}(t) = \cos^{-1}\left(\cos(\theta(t))\cos(\varphi(t) + \beta(t))\right)$ is the thrust vector deviation from $Z_I$ and $a_{\max} = \Delta v_d / T_b$, $F_T = a_{\max}(m_s + m_n)$. $\Delta v_d$ is desired velocity change, and $T_b$ is burning time for an orbital transfer mission.

### 2.4 Linearized State Space Model

In order to linearize the nonlinear model (**2**) and (**6**), some assumptions on the parameters are chosen as:

$$\boldsymbol{F}_T^n = \begin{bmatrix} 0 & 0 & F_T \end{bmatrix}^T,\ \boldsymbol{\rho}_{no}^n = \begin{bmatrix} 0 & 0 & z_n \end{bmatrix}^T,\ \boldsymbol{\rho}_{nT}^n = \boldsymbol{0}_{3\times1},$$
$$\boldsymbol{\rho}_{so} = \begin{bmatrix} x_s & y_s & -z_s \end{bmatrix}^T,\ \boldsymbol{I}_n^n = \mathrm{diag}(I_{n2}, I_{n2}, I_{n1}), \tag{8}$$
$$\boldsymbol{I}_s = \mathrm{diag}(I_{s2}, I_{s2}, I_{s1}),\ \boldsymbol{\omega}_s(0) = \begin{bmatrix} 0 & 0 & \bar{\omega}_s \end{bmatrix}^T.$$

Where $z_s$ and $z_n$ are the distance of the pivot o from the C.Ms of the body and the nozzle. Furthermore, $x_s$ and $y_s$ indicate the C.M offsets of the body. $I_{s2}$ and $I_{n2}$ denote the transverse moments of inertia of the body and nozzle, respectively. $I_{s1}$ and $I_{n1}$ are the moment of inertia about the axes $z_s$ and $z_n$. $\bar{\omega}_s$ is the spacecraft initial spin-rate about its longitudinal axis.

From the dynamics (**2**), the disturbance created by the C.M offsets $x_s$ and $y_s$ is obtained as:

$$\bar{\tau}_{dx} = (F_T M y_s)/m_n,\ \bar{\tau}_{dy} = -(F_T M x_s)/m_n, \tag{9}$$

The state vector that should be stabilized is chosen as: $\boldsymbol{X} = [\varphi\ \theta\ \omega_{sx}\ \omega_{sy}\ \beta\ \dot{\beta}]^T$.

Now by considering a small variation for the states of $\boldsymbol{X}$ and using Eq. (**2**), Eq. (**6**), and assumptions in (8), the extended state space model is obtained as following:

$$\dot{\boldsymbol{X}}(t) = \boldsymbol{A}\boldsymbol{X}(t) + \boldsymbol{B}u + \boldsymbol{D}(t), \tag{10}$$

where the voltage $u = V$ is the control input,

www.engconfs.ir 4

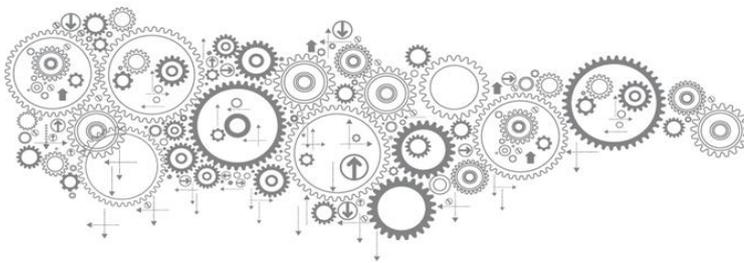

$$A = \begin{bmatrix} 0 & \bar{\omega}_s & 1 & 0 & 0 & 0 \\ -\bar{\omega}_s & 0 & 0 & 1 & 0 & 0 \\ 0 & 0 & 0 & I_r a_{m2}/B_M - \lambda & I_r a_{m3}/B_M - I_\beta & (I_r N_g^2(c_{f,em} + k_{1,em}k_{2,em}))/B_M \\ 0 & 0 & \lambda & 0 & 0 & I_{nM}\bar{\omega}_s \\ 0 & 0 & 0 & 0 & 0 & 1 \\ 0 & 0 & 0 & -a_{m2}/B_M & -a_{m3}/B_M & -(N_g^2(c_{f,em} + k_{1,em}k_{2,em}))/B_M \end{bmatrix},$$

$$B = \begin{bmatrix} 0 \\ 0 \\ -(I_r k_{2,em} N_g)/B_M \\ 0 \\ 0 \\ (k_{2,em} N_g)/B_M \end{bmatrix}, \quad D(t) = \begin{bmatrix} 0 \\ 0 \\ (\bar{\tau}_{dx} + \tau_{rx}(t) + \bar{\omega}_s h_{ry}(t) + w_x(t))(1/I_2 + I_r d_{m1}/B_M) \\ (\bar{\tau}_{dy} + \tau_{ry}(t) - \bar{\omega}_s h_{rx}(t) + w_y(t))(1/I_2) \\ 0 \\ (\bar{\tau}_{dx} + \tau_{rx}(t) + \bar{\omega}_s h_{ry}(t) + w_x(t))(-d_{m1}/B_M) \end{bmatrix}.$$

$I_1 = I_{n1} + I_{s1}$, $I_2 = I_{n2} + I_{s2} + M(z_n + z_s)^2$, $(I_1 - I_2)\bar{\omega}_s = I_2 \lambda$,

$I_{n2} + M(z_n + z_s)z_n = I_r I_2$, $I_{n1} - I_{n2} - M(z_n + z_s)z_n = I_{nz} I_2$,

$(F_T M z_s)/m_n + I_{nz} I_2 \bar{\omega}_s^2 = I_\beta I_2$,

$I_{nm} = I_{nz} - I_r \Leftrightarrow (I_{n1} - 2I_{n2} - 2M(z_n + z_s)z_n) = I_{nm} I_2$,

and

$a_{m2} = I_{s2}\lambda + (I_{s2} - I_{s1})\bar{\omega}_s + Mz_s(\lambda + \bar{\omega}_s)(z_n + z_s)$,

$a_{m3} = I_\beta I_{s2} + Mz_s(z_n\bar{\omega}_s^2 + I_\beta(z_n + z_s)) - (F_T M z_s)/m_n$,

$B_M = I_r I_{s2} - Mz_s(z_n - I_r(z_n + z_s)) + J_{em}N_g^2$,

$d_{m1} = 1 - (Mz_s^2 + Mz_n z_s + I_{s2})/I_2$.

## 3. CLOSED-LOOP CONTROL SYSTEM DESIGN

For stabilizing the partial states *X*, a full state feedback controller based on gimbal actuator is required. A feed-forward controller based on the two RWs is needed to reject the disturbances in two axes. These two controllers are introduced in the following section.

### 3.1 Gimbaled-Thruster Based Full State Feedback Controller

In [2], the controllability of the mentioned spacecraft equipped with a 1DoF gimbaled-thruster is shown. Then the following linear controller can guarantee the stability of the system:

$u(t) = -KX(t)$, $K \in R^{1 \times n}$ (11)

By using a proper gain *K*, $\tilde{A} = A - BK$ is stable.

### 3.2 RW Based Feed-Forward Controller

In order to reject the disturbance torques, a feed-forward controller based on the two RWs is proposed [2]. By employing the following control rule, the RWs gyroscopic torques can reject the constant disturbances $\bar{\tau}_{dx}$ and $\bar{\tau}_{dy}$.

$\tau_{rx} = \tau_{Rm} \tanh(\gamma e_{rx})$, $\tau_{ry} = \tau_{Rm} \tanh(\gamma e_{ry})$, (12)

Where, $\tau_{Rm}$ denotes the maximum reaction torque of the RWs, $e_{rx}(t) = h_{rx}(t) - \bar{\tau}_{dy}/\bar{\omega}_s$ and $e_{ry}(t) = h_{ry}(t) + \bar{\tau}_{dx}/\bar{\omega}_s$ are the angular momentum tracking errors and $\gamma$ is a positive gain. In [2], a disturbance observer is proposed in which the estimated values of $\bar{\tau}_{dx}$ and $\bar{\tau}_{dy}$ can be used in the above control logic.

## 4. SIMULATION RESULTS

In Table 1, the specifications of some industrial DC motors are listed. As reported in row 5, DC motors which are designed by *Bosch* company equipped with a gearbox.

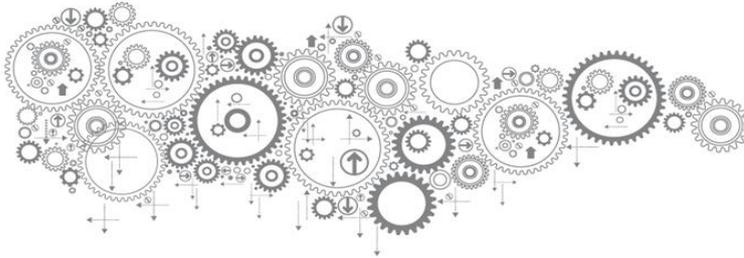

Table 1. Specifications of some suitable industrial DC motors.

| Case | Model | Mass (kg) | $\tau_{em,m}$ (Nm) | $\omega_{em,m}$ (rad/s) | $V_{em,m}$ (V) |
|---|---|---|---|---|---|
| 1 | FAULHABER (3272_CR_DFF) | 0.32 | 1.2 | 540 | 12 |
| 2 | FAULHABER (3890_CR_DFF) | 0.55 | 2.65 | 540 | 18 |
| 3 | buehler motor (EC-Motor_39x107_1.25.037.4XX) | 0.5 | 1.05 | 450 | 12 |
| 4 | buehler motor, (EC-Motor_62x112_1.25.058.2XX) | 1.4 | 3.8 | 400 | 24 |
| 5 | bosch-ibusiness (F 006 B20 092) | 1 | 27 | 4.4 | 12 |
| 6 | maxon motor | 0.48 | 1.72 | 692 | 12 |

For carrying out the simulations, the parameters are selected as: $\bar{\omega}_s = 6$ rad/s, $N_g$=10, disturbance torque is 12.64Nm (4cm thrust misalignment from the C.M), and $\tau_{Rm} = 0.2$ Nm. The DC motor "FAULHABER 3890_CR_DFF" is chosen with the specifications of mass 0.55kg, maximum voltage 18v, stall torque 2.65Nm, stall current 79A, maximum speed 540rad/s, $J_{em} = 164 \, \text{gcm}^2$, and $\tau_{f,em} = 10$ mNm. The control input is $u = V = -KX$ with the following control gain:
$K = [31.82 \quad -131.44 \quad -65.24 \quad 18.32 \quad 217.39 \quad -0.37]$.
The other parameters are
$\Delta v_d = 100 \, m/s$, $T_b = 50s$, $I_{s2} = 10 kgm^2$, $I_{s1} = 2.4 I_{s2}$,
$I_{n2} = 1 kgm^2$, $I_{n1} = 0.5 I_{n2}$,
$m_s = 150 kg$, $m_s = 8 kg$, $z_n = 0.2m$, $z_s = 0.75m$.
The maximum acceleration and thrust force become $a_{max}$=2m/s$^2$ and $F_T$=($m_n$+$m_s$) $a_{max}$=316N, respectively. $\delta_{FT,m}$ and $\bar{\delta}_{FT}$ are also calculated in this simulation where they represent the maximum and mean values of thrust vector devotion, respectively.

Spacecraft body attitude ($\varphi, \theta$) and thrust vector deviation ($\delta_{FT}$) are shown in Figure 5. $\delta_{FT}$ is fully eliminated with the maximum overshoot of $\delta_{FT,m} = 9.43^0$ and the average value of $\bar{\delta}_{FT} = 1.16^0$, as presented in Table 2 By passing the time, the disturbances are fully rejected by the gyroscopic effect of the RWs. An accurate velocity change, $v_z$=99.88m/s, is achieved in comparison with $\Delta v_d$=100m/s. The gimbal angle β and its rate $\dot{β}$, are presented in Figure 6. The maximum deflection of the gimbal angle is 6.34deg. As shown in Figure 7, the RWs are activated by their maximum reaction torques ($\tau_{Rm} = 0.2 Nm$) to reject the disturbances quickly.

Table 2. Maximum overshoots of the state variables.

| $\varphi_m$ (deg) | $\theta_m$ (deg) | $\beta_m$ (deg) | $\dot{\beta}_m$ (deg/s) | $u_m$ (rad/s$^2$) | $\delta_{FT,m}$ (deg) | $v_z$ (m/s) |
|---|---|---|---|---|---|---|
| 4.00 | 5.71 | 6.34 | 13.45 | 1.84 | 9.43 | 99.88 |

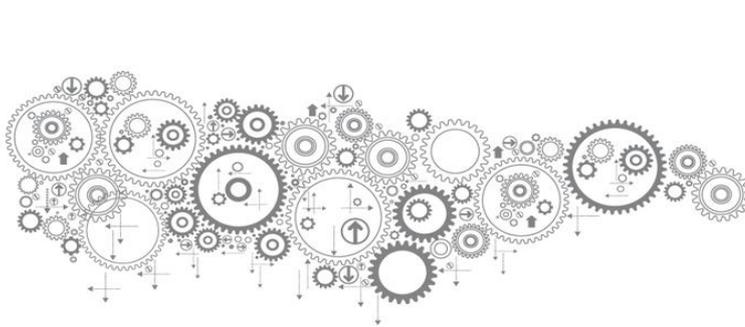

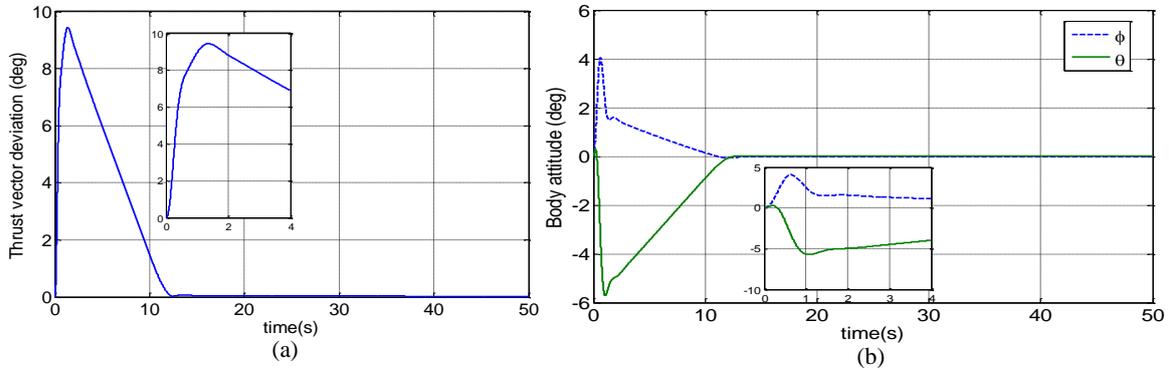

**Figure 5.** Time variation of (a) thrust vector deviation ($\delta_{FT}$), and (b) body attitude ($\varphi, \theta$)

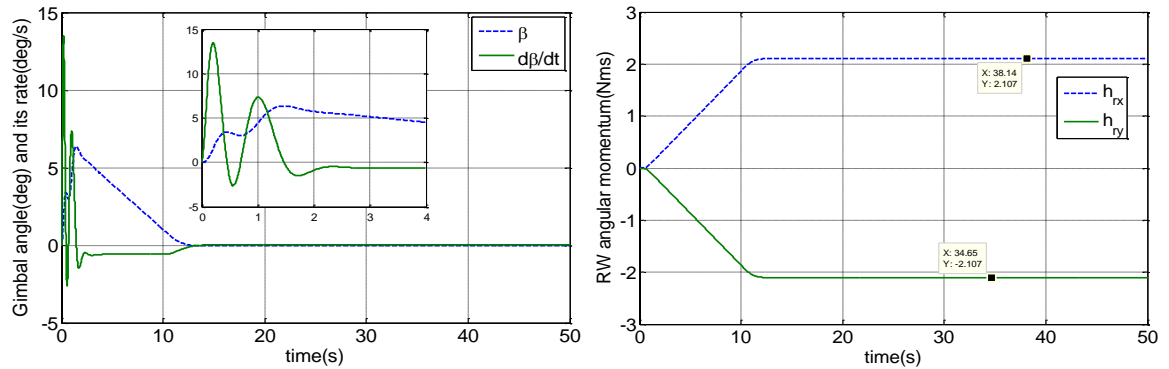

**Figure 6.** Gimbal angle and its rate ($\beta, \dot{\beta}$)    **Figure 7.** The RWs angular momentum

In Figure 8, the torque and speed of the gimbal and DC motor are indicated. In Figure 9, the motor current and its input voltage are given. In Table 3, the overshoot of the DC motor variables is presented that is so smaller than the maximum values presented in Table 1. Therefore, the applicability of the proposed control method is confirmed by employing an industrial DC motor.

**Table 3.** Maximum overshoot of the DC motor variables.

| $V_{em,m}$ (v) | $\omega_{em,\max}$ (rad/s) | $\tau_{em,m}$ (Nm) | $I_{em,m}$ (A) |
|---|---|---|---|
| 4.37 | 2.34 | 0.66 | 19.86 |

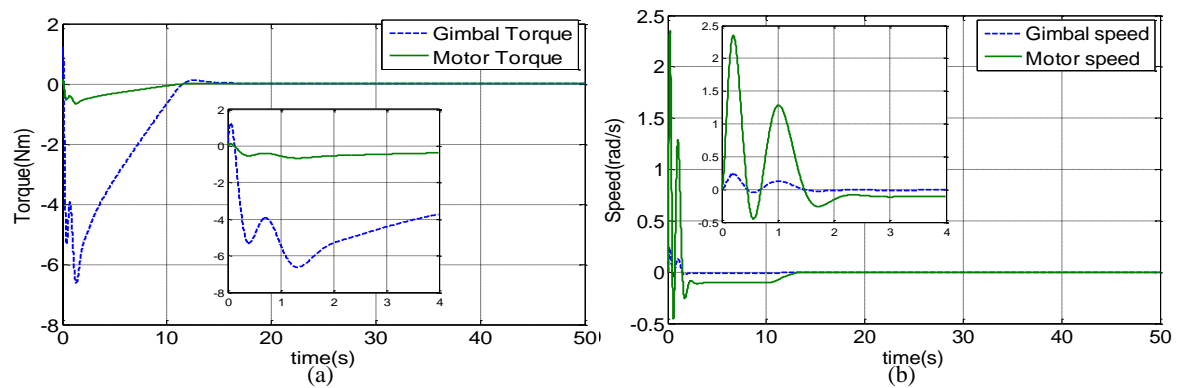

**Figure 8.** Time variation of (a) torque, (b) and speed of the gimbal and DC motor

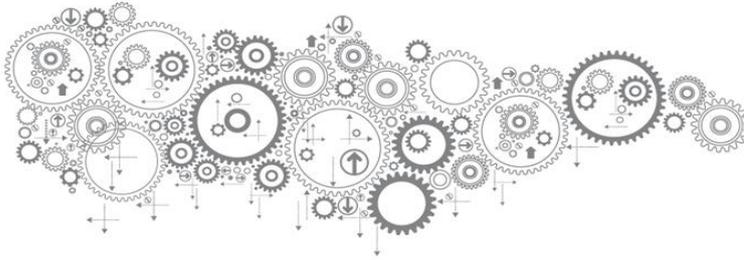

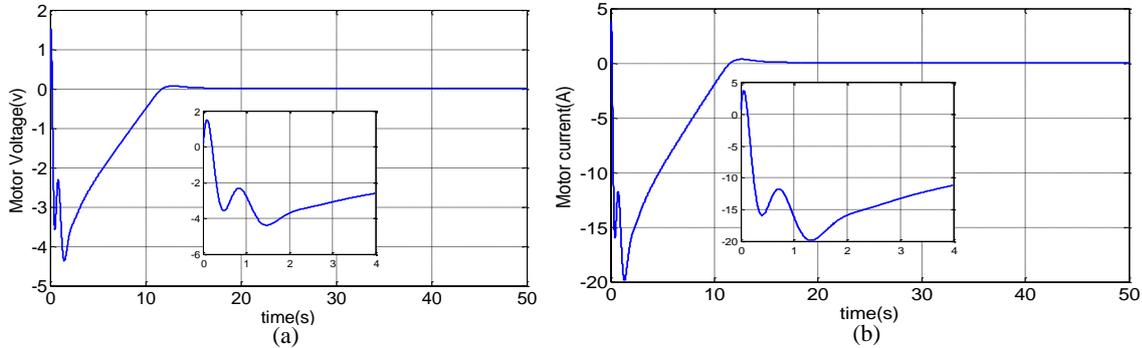

**Figure 9.** The motor (a) input voltage, (b) and its current

## 5. CONCLUSION

The control method based on spin-stabilization, a 1DoF gimbaled-thruster, and two RWs is an important and novel control method. Thus, the ability of the gimbal actuator for stabilizing the mentioned spacecraft is investigated in this paper. To achieve this goal, a DC electric motor with a gearbox is considered for the gimbal actuator. In this work, the spacecraft dynamics model is extended by considering the DC motor equations. To check the control system's applicability, a real industrial DC motor is selected for the gimbal actuator; then, numerical simulation is carried out according to its specifications. The results proved that the control system equipped with a DC motor could stabilize the spacecraft attitude as well as reject the disturbance torque. The good performance of the aforementioned spacecraft is shown for deploying in an impulsive orbital maneuver.

**Nomenclature**

| | |
|---|---|
| $m_n$ | nozzle mass, kg |
| $K$ | linear controller gain |
| $F_T$ | thrust force, N |
| $G_n$ | nozzle C.M location |
| $G_s$ | body C.M location |
| $\Delta v_d$ | desired velocity change, m/s |
| $x_s y_s z_s$ | body-fixed coordinate frame |
| $\Delta v$ | velocity change increment, m/s |
| $F_o$ | interaction force at the pivot o |
| $V_{em}$ | Motor voltage, V |
| $N_g$ | ration of gearbox |
| $k_{1,em}$ | the motor constants |
| $m_s$ | body mass, kg |
| u | control input, V |
| $I_{n2}$ | nozzle moment of inertia, kgm$^2$ |
| $I_{s2}$ | body moment of inertia, kgm$^2$ |
| $T_b$ | burning time , s |
| $X_I Y_I Z_I$ | inertial coordinate system |
| $x_n y_n z_n$ | nozzle-fixed coordinate frame |
| $M_o$ | interaction torque at the pivot o, Nm |
| $H_{Rw}$ | RWs angular momentum, kg m$^2$/s |
| $z_n$ | the distance of the pivot o from the C.Ms of the nozzle, m |

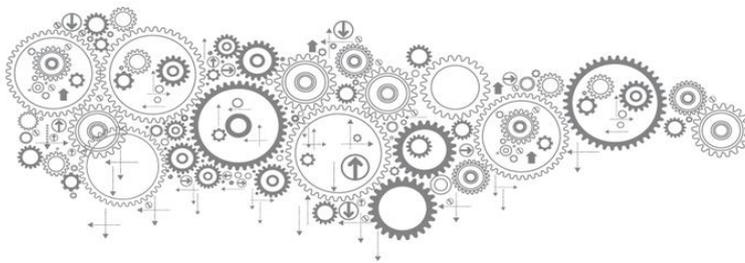

| | |
|---|---|
| $z_s$ | the distance of the pivot o from the C.Ms of the body, m |

**Greek symbols**

| | |
|---|---|
| $\beta$ | Gimbal rotation angle, rad |
| $\theta$ | body attitude angle, rad |
| $\tau_s$ | body external torque, Nm |
| $\tau_n$ | nozzle external torque, Nm |
| $\tau_d$ | exogenous disturbance, Nm |
| $\omega_n$ | nozzle angular velocity, rad/s |
| $\tau_{Rw}$ | RWs axial torque, Nm |
| $\tau_{em}$ | the output torque, Nm |
| $\tau_{f,em}$ | the friction torque, Nm |
| $\delta_{FT}$ | thrust vector deviation from the $Z_I$, rad |
| $\dot{\omega}_r$ | angular acceleration of the nozzle with respect to the body, rad/s$^2$ |
| $\omega_r$ | angular velocity of the nozzle with respect to the body, rad/s |
| $\dot{\omega}_s$ | spacecraft body angular acceleration, rad/s$^2$ |
| $\omega_s$ | body angular velocity, rad/s |
| $\dot{\omega}_{em}$ | the angular acceleration of the rotor, rad/s |
| $\omega_{em}$ | the angular velocity of the rotor, rad/s |

## REFERENCES

[1] Orr JS, Shtessel YB (2012), Lunar spacecraft powered descent control using higher-order sliding mode techniques, *Journal of the Franklin Institute* 349:476-492 doi:http://dx.doi.org/10.1016/j.jfranklin.2011.06.015.

[2] Kouhi H, Kabganian M, Saberi FF, Shahravi M (2017b), Robust control of a spin-stabilized spacecraft via a 1DoF gimbaled-thruster and two reaction wheels, *ISA transactions* 66:310-324

[3] Felicetti L, Sabatini M, Pisculli A, Gasbarri P, Palmerini GB Adaptive Thrust Vector Control during On-Orbit Servicing. In: Proceedings of AIAA SPACE 2014 Conference and Exposition, paper *AIAA-2014-4341, San Diego*, 2014.

[4] Rizvi F, Weitl RM (2013), Characterizing Limit Cycles in the Cassini Thrust Vector Control System *Journal of Guidance, Control, and Dynamics,* 36:1490-1500

[5] Reed JMaB (2014) Implementation of the Orbital Maneuvering System Engine and Thrust Vector Control for the *European Service Module*.

[6] Kong F, Jin Y, Kim HD (2016), Thrust vector control of supersonic nozzle flow using a moving plate, *Journal of Mechanical Science and Technology* 30:1209-1216

[7] Noll R (1971), Spacecraft Attitude Control During Thrusting Maneuvers, *NASA* SP-8059

[8] Kouhi H, Kabganian M, Shahravi M, Fani Saberi F (2016). Retrofiring control method via combination of a 1DoF gimbaled thrust vector control and spin-stabilization Proceedings of the Institution of Mechanical Engineers. *Part G: Journal of Aerospace Engineering* doi:10.1177/0954410016650909

[9] Wang Z, Jia Y, Jin L, Duan J (2016). Thrust Vector Control of Upper Stage with a Gimbaled Thruster during Orbit Transfer. *Acta Astronautica.*

[10] Reyhanoglu M, Hervas JR (2012). Nonlinear dynamics and control of space vehicles with multiple fuel slosh modes. *Control Engineering Practice* 20:912-918

[11] Rubio Hervas J, Reyhanoglu M (2014). Thrust-vector control of a three-axis stabilized upper-stage rocket with fuel slosh dynamics. *Acta Astronautica* 98:120-127 doi:http://dx.doi.org/10.1016/j.actaastro.2014.01.022

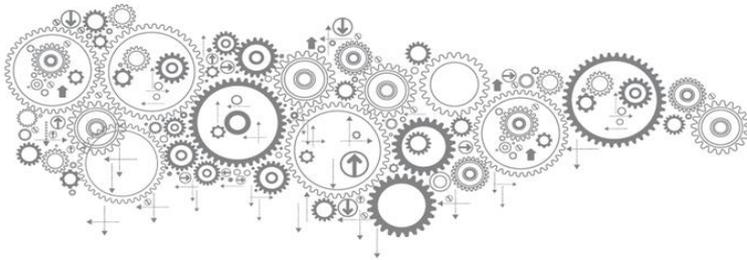
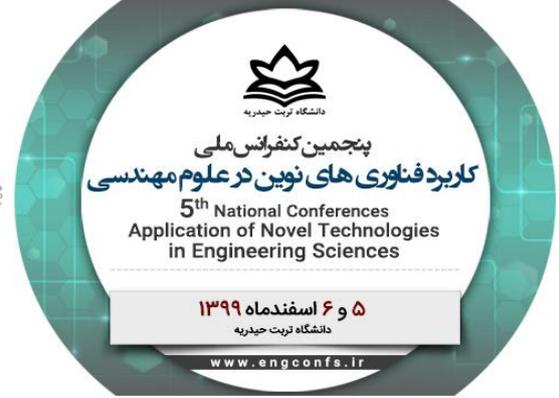

[12]     Hervas JR, Reyhanoglu M (2014). Thrust-vector control of a three-axis stabilized upper-stage rocket with fuel slosh dynamics. *Acta Astronautica* 98:120-127

[13]     Kouhi H, Kabganian M, Fani Saberi F, Shahravi M (2017a). Adaptive control of a spin-stabilized spacecraft using two reaction wheels and a 1DoF gimbaled-thruster. *AUT Journal of Modeling and Simulation* 49:103-112